\documentclass{article}
\usepackage{spconf,amsmath,graphicx,hyperref}
\usepackage{tabularx} 
\usepackage{geometry}
\usepackage{booktabs}
\usepackage{caption}  
\usepackage{graphicx} 
\usepackage{array} 
\usepackage{tablefootnote}
\usepackage{afterpage}
\usepackage{makecell}
\usepackage{placeins}   

\title{Multimodal Privacy-Preserving Entity Resolution with Fully Homomorphic Encryption}
\name{Susim Roy \qquad Nalini Ratha}
  
  \address{University at Buffalo, The State University of New York \\
      Department of Computer Science and Engineering}

\begin{document}
%
\maketitle

\begingroup
\renewcommand\thefootnote{}
\footnotetext{
\textcopyright~ Copyright 2026 IEEE. Published in ICASSP 2026 - 2026 IEEE International Conference on Acoustics, Speech and Signal Processing (ICASSP), scheduled for 3-8 May 2026 in Barcelona, Spain. Personal use of this material is permitted. However, permission to reprint/republish this material for advertising or promotional purposes or for creating new collective works for resale or redistribution to servers or lists, or to reuse any copyrighted component of this work in other works, must be obtained from the IEEE. Contact: Manager, Copyrights and Permissions / IEEE Service Center / 445 Hoes Lane / P.O. Box 1331 / Piscataway, NJ 08855-1331, USA. Telephone: + Intl. 908-562-3966.}
\endgroup
\thispagestyle{empty}
\pagestyle{empty}
\begin{abstract}
The canonical challenge of entity resolution within high-compliance sectors, where secure identity reconciliation is frequently confounded by significant data heterogeneity, including syntactic variations in personal identifiers, is a longstanding and complex problem. To this end, we introduce a novel multimodal framework operating with the voluminous data sets typical of government and financial institutions. Specifically, our methodology is designed to address the tripartite challenge of data volume, matching fidelity, and privacy. Consequently, the underlying plaintext of personally identifiable information remains computationally inaccessible throughout the matching lifecycle, empowering institutions to rigorously satisfy stringent regulatory mandates with cryptographic assurances of client confidentiality while achieving a demonstrably low equal error rate and maintaining computational tractability at scale.
\end{abstract}
\section{Introduction}
\begin{table}[!t]
\centering
\caption{Example dataset entries where Rows 1 and 2 have the same name but different addresses and headshots and Rows 3 and 4 have slight name and address variations with young and aged headshots, respectively.}
\label{tab:data_in_columns}
\setlength{\tabcolsep}{0.5pt}
\renewcommand{\arraystretch}{0.5}
\begin{tabular}{|>{\centering\arraybackslash}m{1.5cm}|
                >{\centering\arraybackslash}m{3.5cm}|
                >{\centering\arraybackslash}m{2cm}|}
\hline
\textbf{Name} & \textbf{Address} & \textbf{Image} \\
\hline
Jonathan Diaz & 8708 Harris Road, Popetown, MH 23945 &
\includegraphics[width=1.5cm]{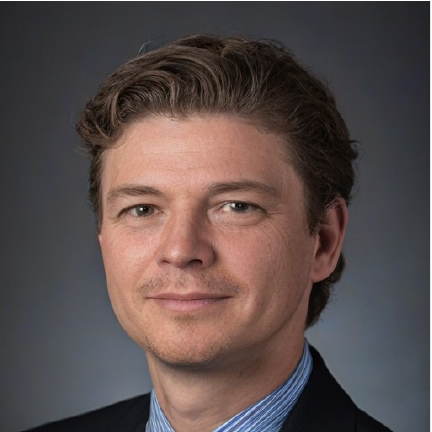} \\
\hline
Jonathan Diaz & 5190 Thomas Mews, Dicksonburgh, TN 13160 &
\includegraphics[width=1.5cm]{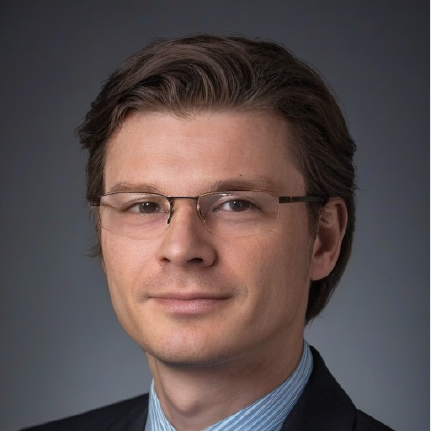} \\
\hline
Mrs. Ann Pham & 254 Savage River Ste. 120, Miguelfort, GA 9209 &
\includegraphics[width=1.5cm]{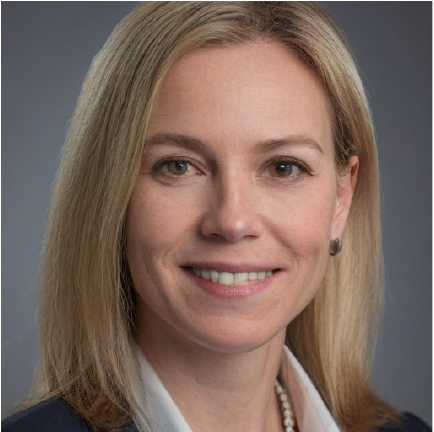} \\
\hline
Anne Pham & 254 Savage River Suite 120, Miguelfort, GA 92091 &
\includegraphics[width=1.5cm]{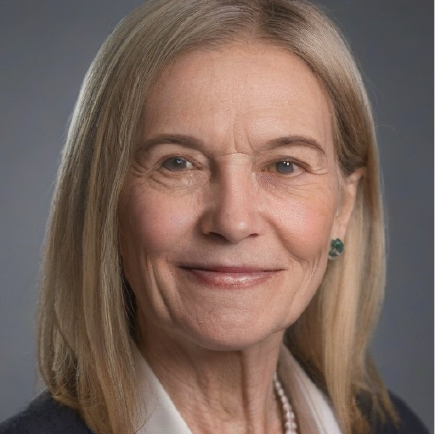} \\
\hline
\end{tabular}
\vspace{-0.8em}
\end{table}
In critical identity verification processes employed by organizations like the Department of Motor Vehicles(DMVs), passport authorities, and financial institutions, a fundamental challenge is reconciling an individual's identity over time. While core biometric identifiers like fingerprints and facial structures change minimally, it may not be available because they were not collected since such modalities were not viable for automated matching or unsuitable for biometric matching due to poor quality. The biographical data is often dynamic; individuals frequently relocate, abbreviate their names, or update addresses. This discrepancy creates significant validation hurdles, especially in scenarios where a person has visibly aged while residing far from their last recorded location. This highlights the critical need for automated systems that can holistically verify an identity before issuing official credentials such as passports and driver's licenses or granting access to regulated services. 

We identify the main culprit for the aforementioned limitations as relying on a single data source while each source may provide significant identity resolution capacity. A natural extension moves toward multimodal data sources. This synergy of combining different data types can produce more reliable and comprehensive authentication outcomes. Existing methodologies \cite{anil_fusion, Bhatt_2013_CVPR_Workshops,nratha,bolme}, while addressing multimodal identity traits for entity resolution, are constrained by limitations such as dependence on unaligned and restricted Pinellas County Sheriff's Office(PCSO) dataset, susceptibility to the inherent recognizability of celebrity faces by Vision-Language Models~\cite{Hintersdorf2022DoesCK}, susceptibility to uncontrolled headshot variations, or reliance on highly specific biographic data. Additionally, utilizing multiple biometric signatures also elevates the security risks. Biometric templates are not just random data; they contain significant information that could potentially be used to reconstruct a person's identity or leak other soft-biometric details. Therefore, it is imperative that any fusion technique is not only robust but also preserves the privacy of the individual's data. To address this critical need, researchers have turned to advanced cryptographic methods. Homomorphic Encryption (HE), for instance, allows for computations to be performed directly on encrypted data. This enables the fusion of biometric templates without ever decrypting them, thus keeping the sensitive information secure throughout the process \cite{sperling2022heft}. This principle extends to the fusion of classifier outputs in ensemble learning systems, where Fully Homomorphic Encryption (FHE) can be used to build privacy-preserving algorithms that securely process data while it remains encrypted, ensuring both high performance and stringent data security\cite{10.1007/978-3-031-78107-0_13}.

In contrast to the previous work, our contribution is as follows: (1) Entity Resolution is processed in an encrypted domain preserving the user's privacy (2) Multimodal learning stream using \textbf{biometric} and \textbf{biographic} data (3) We release a novel synthetic dataset simulating a real-world entity resolution scenario.
\vspace{-10pt}
\section{Methodology}
\begin{figure}[t]
    \centering
    \includegraphics[width=\linewidth]{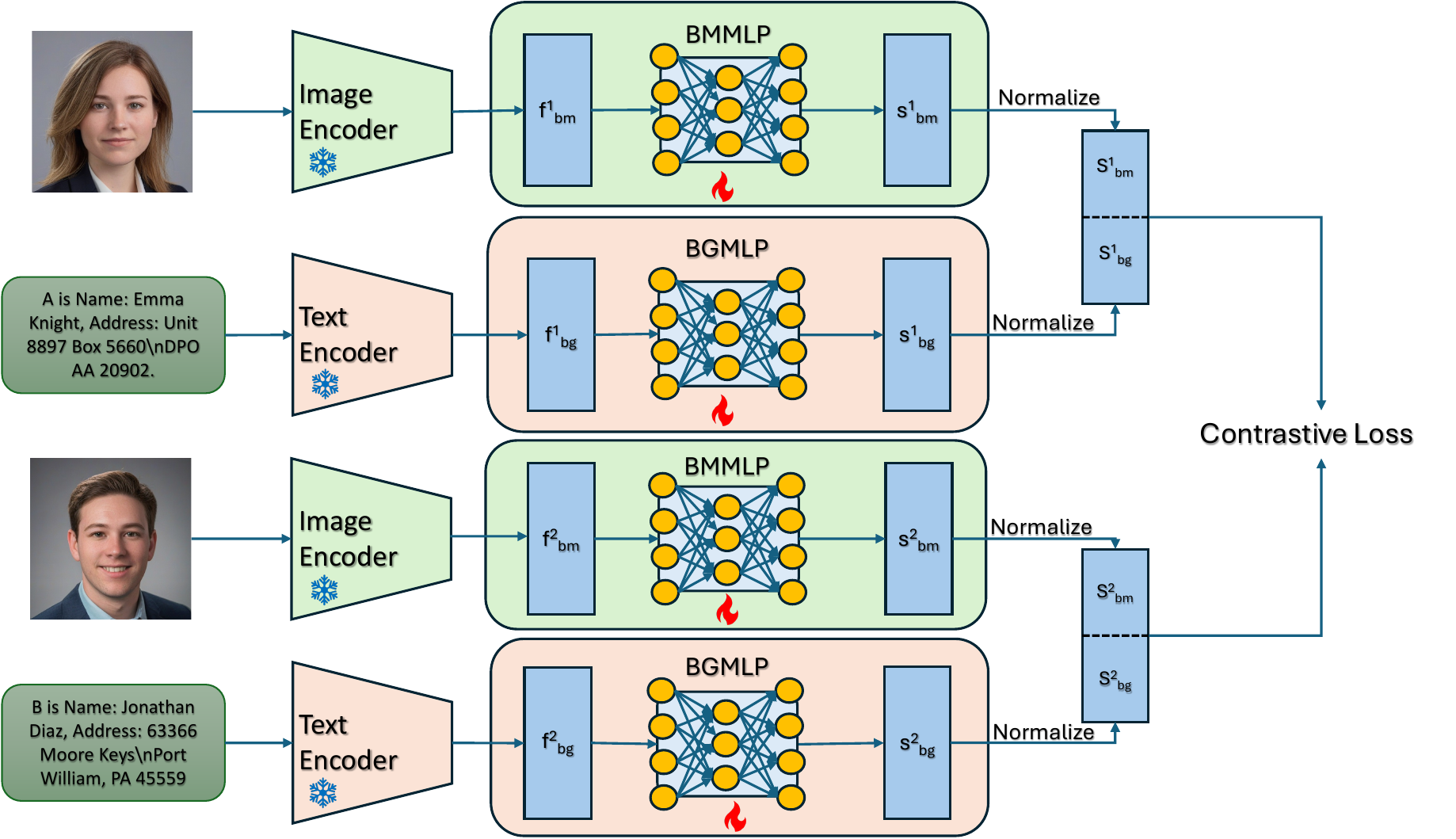}
    \caption{The above figure demonstrates the training pipeline to obtain the biometric and biographic score vectors. Here, $s^1_{bm}$ and $s^1_{bg}$ denote the score vectors for one unique individual.}
    \label{fig:training}
    \vspace*{-1.2em}
\end{figure}
\subsection{Dataset and Pipeline Overview}
\begin{figure*}[t]
    \centering
    \includegraphics[width=0.72\textwidth, keepaspectratio]{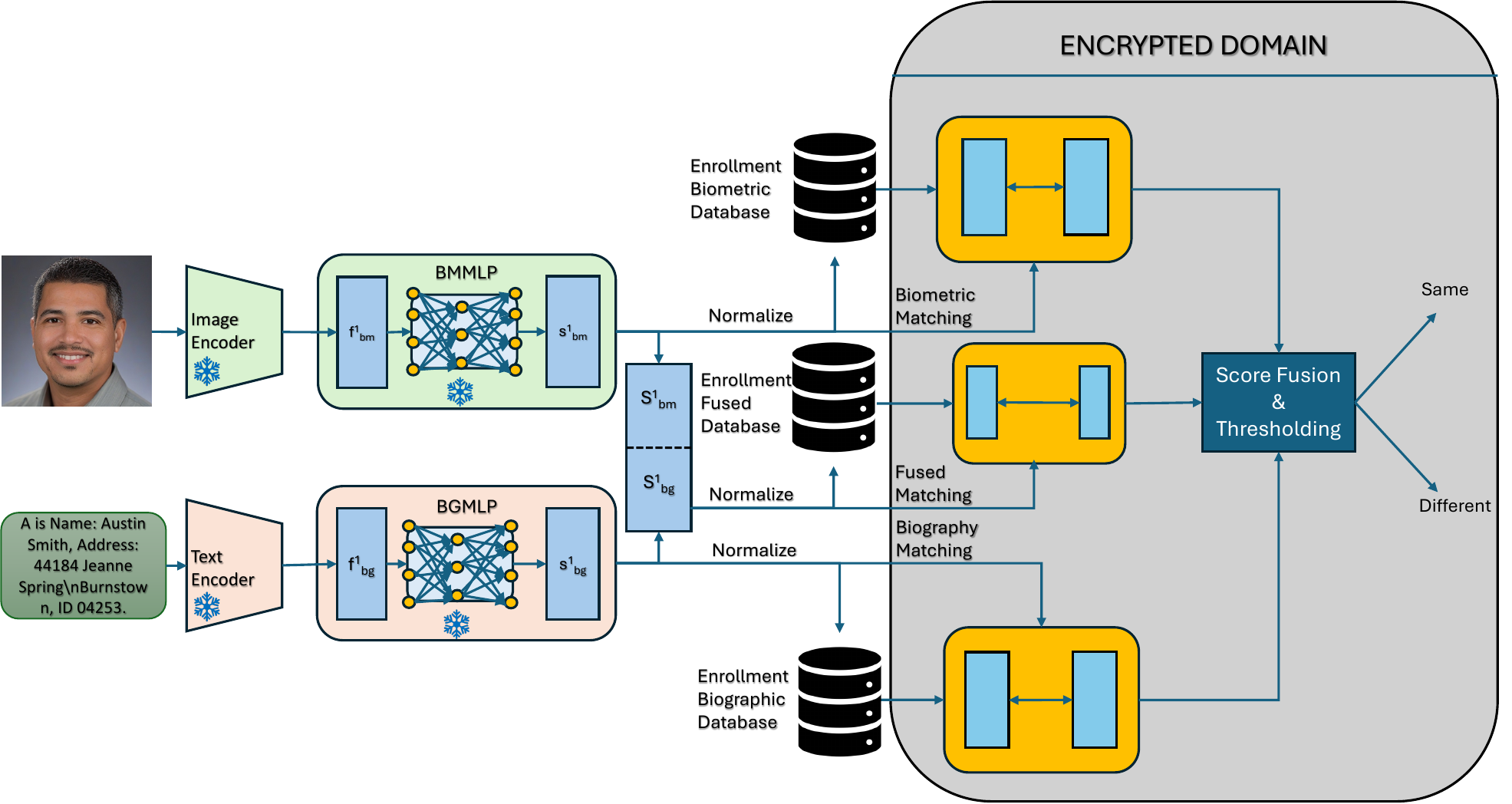}
    \caption{The above figure demonstrates the encrypted enrollment and authentication pipeline. The three modality databases store the encrypted biometric, fused, and biographic templates, which are then matched during verification.}
    \label{fig:testing}
\end{figure*}
\begin{table*}[h!t]
\centering
\caption{The table below shows the summary of our multimodal dataset creation and statistics. The alternate address spellings and common name variations were created with Claude 3.5-Sonnet\cite{anthropic2024claude35sonnet} based on the LM-Arena rankings.}
\label{tab:dataset_stats}
\begin{tabular}{@{}llrrr@{}}
\toprule
\textbf{Stage} & \textbf{Description} & \textbf{\# Individuals} & \textbf{\# Records} & \textbf{\# Images} \\
\midrule
\multicolumn{5}{l}{\textit{Data Augmentation Steps}} \\
Initial & Base unique individuals & 36,661 & 36,661 & 36,661 \\
Step 1 & Added alternate address spellings & 36,661 & 46,365 & 46,365 \\
Step 2 & Added 4-7 random new addresses per person & 36,661 & 266,098 & 266,098 \\
Step 3 & Added common name variations with 2-7 addresses & 36,661 & 287,532 & 287,532 \\
\midrule
\multicolumn{5}{l}{\textit{Final Dataset Split}} \\
Training Set & Used for model training & 21,661 & $\sim$277,532 & $\sim$277,532 \\
Testing Set & Used for evaluation & 10,000 & 10,000 & 10,000 \\
- Enrolled & Individuals registered in the test database & 5,000 & 5,000 & 5,000 \\
- Verification & Individuals used for test queries & 5,000 & 5,000 & 5,000 \\
\midrule
\textbf{Total} & \textbf{Final Generated Dataset} & \textbf{36,661} & \textbf{287,532} & \textbf{287,532} \\
\bottomrule
\end{tabular}
\end{table*}
We build on top of \cite{mishra2024synthetic} by selecting the $\langle \text{Name}, \text{Address} \rangle$ attributes out of 8 attributes related to the financial information of different synthetically generated people. Now let us consider the following scenarios where an individual's biographic data maybe altered (1) An individual can write their name or address in different formats either intentionally or mistakenly, e.g., \textit{Anne}, \textit{Street} and \textit{Boulevard} can be written as \textit{Ann}, \textit{St.} and \textit{Blvd} respectively(2) An individual may have lived at different addresses overlapping with another individuals's addresses (3) Each individual's biographic data might be related to different headshot(biometric) images taken over different time periods, e.g., a person might look aged than when they first did their Know Your Customer(KYC). We used Stable Diffusion XL\cite{conf/iclr/PodellELBDMPR24} to first create a base image on top of which the variegated headshots were produced. Table \ref{tab:dataset_stats} shows the detailed description of the number of datapoints in our synthetically created dataset after each step and the final set. An example of the data points is shown in Table \ref{tab:data_in_columns}. \\
On the training set, we first train our model for $5$ epochs with a learning rate of $1e-5$ and batch size of $256$ as shown in Figure \ref{fig:training}. This trained model is used to store 5k entity templates in their respective databases. Once stored, they are then used for entity resolution with query and gallery templates in an encrypted space.
\vspace{-5pt}
\subsection{Training Procedure}
For training, we use the image and text encoders from the CLIP ViT-B/32 model\cite{Radford2021LearningTV}. We use pretrained weights to obtain the biometric template($f^i_{bm}$) from the image encoder and the biographic template ($f^i_{bg}$) from the text encoder. The biometric template and the biographic templates are then processed by two separate trainable modules, BMMLP and BGMLP, as demonstrated in Figure \ref{fig:training}. BMMLP produces the biometric score vector$(s^i_{bm})$ and BGMLP produces the biographic score vector$(s^i_{bg})$. These vectors are then normalized and concatenated and trained via contrastive loss\cite{1640964}.\\
Let us consider that we have a set of $n$ biometric templates. To bring $k$ out of $n$ biometric templates about the same individual closer, the BMMLP module share their weights across the set of template vectors $\{f^1_{bm},f^2_{bm},\dots,f^n_{bm}\}$ to produce the set of score vectors $\{s^1_{bm},s^2_{bm},\dots,s^n_{bm}\}$, same as the BGMLP module. For biographic score vectors, we follow \cite{tang2022generic} to make the prompt where an entity $a$ is represented with a list of attribute-value pairs $(att_i, val_i)_{1\leq i \leq k}$ which will be tokenized as:
\begin{equation*}
\begin{split}
 \text{tokenize}(a) = [ATT]\text{att}_1[VAL]\text{val}_1 \dots \\
    [ATT]\text{att}_k[VAL]\text{val}_k
\end{split}
\end{equation*}
\vspace{-10pt}
\subsection{Encrypted Enrollment and Verification}
Figure \ref{fig:testing} demonstrates the operations in the ciphertext domain. Enrollment during entity resolution involves generating the normalized score vectors $s^1_{bm}$, $s^1_{bg}$, $\text{concat}(s^1_{bm},s^1_{bg})$ in the respective databases. We follow the Min-Max Normalization, where the minimum and maximum values are found based on the training data. Next, we encrypt the bimodal score vectors using FHE to store them. Specifically, we use the RNS-CKKS scheme from the HEAAN\cite{cryptoeprint:2016/421} library to encrypt them. Once encrypted, we can then match the normalized gallery($g$) and query($q$) vectors with the distance function D defined as $D(q,g) = ||q-g||^2$. This gives us a 1:N set of scores $\{(s'_{1_{bm}},s'_{2_{bm}},\dots,s'_{N_{bm}}),(s'_{1_{bg}},s'_{2_{bg}},\dots,s'_{N_{bg}})\}$ for score-level fusion and  $\{s'_1,s'_2,\dots,s'_N\}$ for feature-level fusion for each query vector $q$.\\
While in theory HE can compute an arbitrary function, its practicality is limited to additions and multiplications. Most non-linear functions need to be approximated with polynomial algorithms. In light of this, we define $comp(a,b) = 0$ if $a<b$; $1$ if $a>b$; and $0.5$ otherwise. Therefore to do this in FHE, we employ the algorithm by \cite{10.1007/978-3-030-34621-8_15} where we start with a variable $x=a-b$ which is transformed by a series of iterative polynomial approximations $\{g_1,\dots,g_n\}$ to efficiently emulate $comp(a,b)$ as a sequence of additions and multiplications, making it amenable to evaluation under FHE. \\
Based on the aforementioned techniques, we first encrypt the vectors($\{s'_{1_{bm}}, s'_{1_{bg}}\}$), normalize using respective min-max normalization ranges, followed by combining them by taking their average. The fused encrypted result is then compared against a pre-determined encrypted threshold based on the training ROC-AUC curve. In the case of feature-level fusion, the encrypted concatenated vector is directly compared under the FHE scheme.
\vspace{-5pt}
\section{Experiments and Results}
\subsection{Robustness of Verification and Identification}
\begin{figure*}[!t]
    \centering
 \includegraphics[height=5.0cm, width=\textwidth]{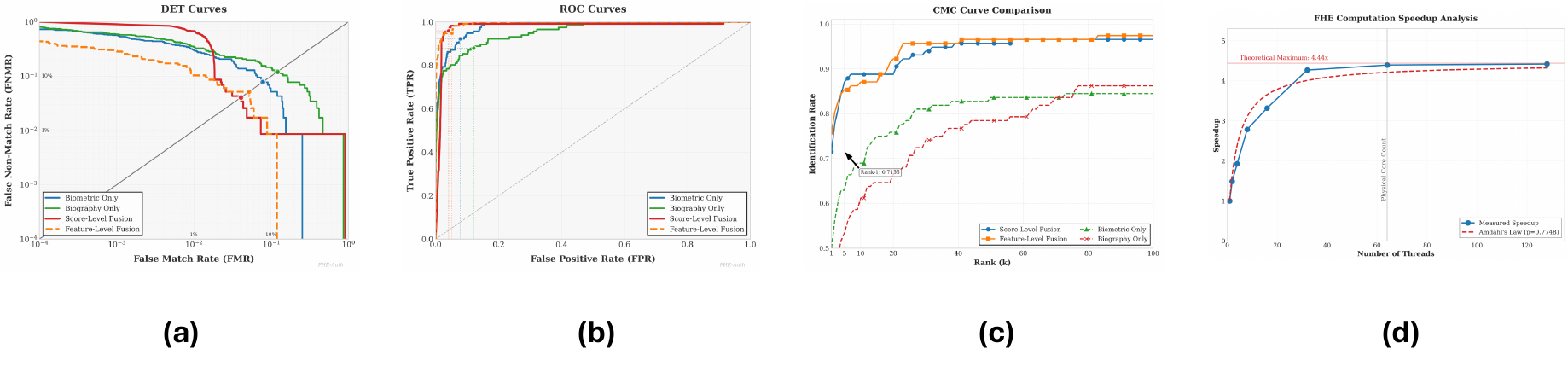}
    \caption{DET(a), ROC(b) and CMC(c) curves comparing verification and identification performance across modalities. The unimodal baselines (biometric and biographic) underperform relative to score-level and feature-level fusion. Figure (d) shows our method asymptotically approaching the theoretical maximum speedup possible with increasing threads. Note that the plaintext results are identical to the corresponding ciphertext results and hence are not shown.}
    \label{fig:roc}
\end{figure*}
\begin{table}[]
\centering
\caption{Ciphertext(CT) matching under CKKS yields identical performance to Plaintext(PT), showing that security is obtained without utility loss.}
\label{tab:results}
\begin{tabular}{@{}|c|c|c|c|@{}}
\toprule
\textbf{Method} & \textbf{EER(\%)} & \multicolumn{2}{c|}{\textbf{TPR@FMR}} \\ \cmidrule(lr){3-4}
       &         & \textbf{1e-2} & \textbf{1e-3} \\ \midrule
Biographic-Only        & 12.37 & 0.646 & 0.396 \\ \midrule
Biometric-Only          & 7.78  & 0.689 & 0.431 \\ \midrule
Fusion (PT-Feature)& 5.17  & 0.896 & 0.706 \\ \midrule
Fusion (PT-Score)        & 4.08  & 0.879 & 0.707 \\ \midrule
Fusion (CT-Feature)& 5.17  & \textbf{0.896} & 0.706 \\ \midrule
Fusion (CT-Score)  & \textbf{4.08}  & 0.879 & \textbf{0.707} \\ \bottomrule
\end{tabular}
\vspace{-1.2em}
\end{table}
Table \ref{tab:results} demonstrates the verification performance of both fusion at the feature level and at the score level in the encrypted domain over unimodal biometric baselines. As observed, the encrypted thresholding not only preserves the plaintext EER but also improves the TPR by 2\% when feature-level encryption is used. Thus, depending on the user's needs, they may choose a particular encryption scheme based on the metric to optimize. \\
We also observe in Figure \ref{fig:roc}(a), (b), and (c), where we notice that the ROC curves show that feature-level and score-level fusion outperform unimodal baselines, achieving higher TPR at the same FPR. Additionally, the DET curves provide a finer view at low FMRs, where bimodal fusion yields the lowest error trajectory. Both plots clearly demonstrate that encryption preserves the entity identification utility gains of score-level and feature-level fusion without degrading performance. We also observe that the CMC curves reveal significant insights into the identification capability of our authentication system under different fusion strategies. Feature-level fusion demonstrates superior performance at rank-1 identification with an accuracy of \textbf{75.86\%}, outperforming score-level fusion (71.55\%) by over 4\%. This indicates that combining modalities earlier in the processing pipeline preserves more discriminative information. However, an interesting pattern emerges at higher ranks: score-level fusion exhibits better rank-5 accuracy (87.07\% vs. 85.34\%), suggesting that it may better distribute potential matches across nearby ranks when the top match is incorrect. This crossover behavior between the two fusion methods illustrates a trade-off: feature-level fusion optimizes for precise first-match identification, while score-level fusion provides more robust performance when considering multiple potential matches. Both fusion strategies significantly outperform the individual biometric and biographical modalities across all ranks, confirming the effectiveness of multimodal approaches in FHE-based authentication systems.
\subsection{Latency Analysis of FHE Operator}
Our FHE authentication operator demonstrates impressive parallelization efficiency with actual measurements showing a 4.42× speedup with just 128 threads compared to the single-threaded baseline (reducing elapsed time from 20,212 ms to 4,573 ms) as shown in Figure \ref{fig:roc} (d). The performance scaling exhibits classic diminishing returns, with near-linear improvements up to 32 threads before memory bandwidth limitations begin to constrain further gains. While CPU time increases with thread count (from 20.2 ms to 186.5 ms), this reflects the expected overhead of thread management and cache coherence protocols. The measured \textbf{9.22×} speedup when moving from 1 to 64 threads (elapsed time dropping from 20,212 ms to 4,600 ms) demonstrates the effectiveness of our parallel implementation in significantly reducing authentication latency while preserving the security guarantees of fully homomorphic encryption.
\vspace{-12pt}
\section{Conclusion}
\vspace{-10pt}
In this paper, we provide a privacy-preserving method for entity resolution using multimodal data. Our formulation shows that we can achieve high efficiency gains with minimal computation while protecting the user's privacy. As future work, this approach can be extended and evaluated on a larger dataset.
\bibliographystyle{IEEEbib}
\bibliography{strings}

\end{document}